\documentclass[12pt]{article}
\usepackage{color} %
\usepackage{a41} %
\usepackage{float} %
\usepackage{lscape} %
\usepackage{epsfig} %
\usepackage{amssymb} %
\usepackage{amsmath} %
\usepackage{verbatim} %
\usepackage{cite} %
\newcommand{\beq}{\begin{equation}}
\newcommand{\eeq}{\end{equation}}
\newcommand{\bea}{\begin{eqnarray}}
\newcommand{\eea}{\end{eqnarray}}

\newcommand{\lsim}{\raisebox{-0.07cm}{$\, \stackrel{<}{{\scriptstyle
\sim}}\, $}}

\graphicspath{/usr1/hboett/plots/qcd_pol/}
\graphicspath{./}

\begin{document}
\begin{titlepage}

\begin{flushleft}
DESY 11--062  
\\ 
DO--TH 11/11  \\
SFB/CPP--11--18\\
LPN-11--17\\
April 2011 \\
\end{flushleft}
\vspace{1.5cm}
\noindent
\begin{center}
{\LARGE\bf New Exclusion Limits for Dark Gauge Forces} 

\vspace{2mm}
{\LARGE\bf from Beam-Dump Data}
\end{center}
\begin{center}

\vspace{2cm}

{\large Johannes Bl\"umlein$^1$ and J\"urgen Brunner$^{1,2}$}

\vspace{1.5cm}
{\it $^1$~Deutsches Elektronen--Synchrotron, DESY,\\
Platanenallee 6, D--15738 Zeuthen, Germany}

\vspace{4mm}
{\it $^2$~CPPM, Aix-Marseille Universit\'{e}, CNRS/IN2P3, Marseille, France}

\vspace{2.5cm}
\end{center}

\begin{abstract}
\noindent
We re-analyze proton beam dump data taken at the U70 accelerator at IHEP Serpukhov 
with the $\nu$-calorimeter I experiment in 1989 to set mass-coupling limits for 
dark gauge forces. The corresponding data have been used for axion and light Higgs 
particle searches in Refs.~\cite{Blumlein:1990ay,Blumlein:1991xh} before. We determine 
new mass and coupling exclusion bounds for dark gauge bosons. 
\end{abstract}
\end{titlepage}

\vfill
\newpage
\sloppy

\section{Introduction}

\vspace*{1mm}
\noindent
Long range forces based on a $U(1)$ gauge symmetry beyond those of the $SU(3)_c 
\times SU(2)_L \times U(1)_Y$ Standard Model may exist yet unnoticed if their 
coupling to ordinary matter is very weak \cite{Holdom:1985ag}. Symmetries of this
kind are discussed in various extensions of the Standard Model, see the 
surveys~\cite{Bjorken:2009mm,Redondo:2010dp,Andreas:2010tp}. A new $U(1)$ gauge 
boson $\gamma'$ with masses $m_{\gamma'}$ in the MeV--GeV range extends the Lagrangian 
of the Standard Model ${\cal L}_{\rm SM}$ to \cite{Bjorken:2009mm,Essig:2010gu} 
\begin{eqnarray}
\label{eq:1}
{\cal L} = {\cal L}_{\rm SM} 
- \frac{1}{4} X_{\mu\nu} X^{\mu\nu} 
+ \frac{\epsilon}{2} X_{\mu\nu} F^{\mu\nu} + e_\psi \epsilon \overline{\psi} 
\gamma_\mu \psi X^\mu
+ \frac{m_{\gamma'}^2}{2} X_{\mu} X^{\mu}~.
\end{eqnarray}
Here $X^{\mu}$ denotes the new vector potential and $X^{\mu\nu} = \partial^\mu 
X^{\nu} - \partial^\nu X^{\mu}$ the corresponding field strength tensor, with
$F^{\mu\nu}$ the $U(1)_Y$ field strength tensor. The mixing of the new $U(1)$ and 
$U(1)_Y$ of the Standard Model is induced by loops of heavy particles coupling to 
both fields \cite{Holdom:1985ag,Andreas:2010tp}. The field $X_{\mu}$ is assumed to 
couple minimally to all charged Standard Model fermions $\psi$, with effective 
charge $e_\psi \epsilon$, where $e_\psi$ is the fermionic charge under $U(1)_Y$. For 
the generation of the mass term we assume the Stueckelberg formalism 
\cite{Stueckelberg:1900zz}, as an example. Possible other mechanisms consist in
technicolor or spontaneous symmetry breaking. The latter ones would lead to more 
terms in (\ref{eq:1}). The parameter $\epsilon$ denotes the mixing parameter of 
the two $U(1)$ groups and may take values in the range $\epsilon \sim 10^{-23} - 
10^{-2}$, depending on the respective model, cf.~\cite{Bjorken:2009mm}. 

Dark $U(1)$ gauge forces contribute to the anomalous magnetic moments of the 
electron 
and muon. Potential signals may be measured from $\Upsilon(3S)$ decays. The $\gamma'$
particles may be created in electron- and proton beam dumps. So far signals of these
particles have not been detected leading to various exclusion bounds in the 
$m_{\gamma'}-\epsilon$ plane in the range of $\epsilon \in [5 \times 10^{-9}, 
10^{-2}]$ and a series of mass regions in $m_{\gamma'} \in [2~m_e, 
3$~GeV],~cf. e.g.~Refs.~\cite{Bjorken:2009mm,Redondo:2010dp,Andreas:2010tp} and 
\cite{MILLI}.

In the present note we derive new exclusion bounds on dark $\gamma'$ bosons using 
proton beam dump data at $p \sim 70$~GeV. These data have been used in the in axion 
\cite{AXION} and light Higgs boson searches, 
cf.~\cite{PROP,Blumlein:1990ay,Blumlein:1991xh} in the past.
We first describe the production process and the experimental facility, and 
then derive new mass and coupling bounds.
\section{The Production Process}

\vspace*{1mm}
\noindent
The abundant production of $\pi^0$ mesons in proton beam dumps leads to a production 
rate of $\gamma'$
\begin{eqnarray}
\sigma(pp \rightarrow \gamma' X) = 2 \epsilon^2 
\left(1 - \frac{m_{\gamma'}^2}{m_{\pi^0}^2}\right)^3
{\rm Br}(\pi^0 \rightarrow \gamma \gamma) \sigma(pp \rightarrow \pi^0 X)~,
\label{pp-tot}
\end{eqnarray}
through $\pi^0 \rightarrow \hat{\gamma} \hat{\gamma}$, with the mixture
$\hat{\gamma} = (\gamma + \epsilon \gamma')/(1+ \epsilon)$ and ${\rm Br}(\pi^0 
\rightarrow \gamma \gamma) = 0.98823 \pm 0.00034$~\cite{Nakamura:2010zzi},
neglecting contributions in higher powers of $\epsilon$. Here the 
phase space factor of the 2--particle decay has been accounted for 
\cite{Batell:2009di}. 
The mass range of the produced $\gamma'$ is limited by $m_{\gamma'} < 
m_{\pi^0} = 134.976$ MeV~\cite{Nakamura:2010zzi}. In principle this range may be 
expanded to higher meson 
decay thresholds including $\eta, \rho^{\pm}, \omega$, and $\eta'$ production, 
since the decay spectra of  these particles contain a large fraction of photons, 
see also \cite{Barabash:1992uq,Essig:2010gu}. 
However, one has to know the corresponding differential meson production spectra 
for $pp$ resp. $pA$ scattering in detail, which to our knowledge have not been 
measured in the energy region under consideration. 

We consider the collision process $p Fe \rightarrow \pi^0 X$ at a momentum
of the incoming proton of $p = 68.6$~GeV. The differential scattering cross 
sections for the reactions $p p \rightarrow \pi^{\pm} X$ for  values of 
$p_\perp \lsim 1$~GeV were measured \cite{Ammosov:1976zj}. One may use
the representation
\begin{eqnarray}
E \frac{d^2 \sigma(pp \rightarrow \pi^0 X)}{d x_F d p_\perp^2} =
\frac{1}{2} \left[E \frac{d^2 \sigma(pp \rightarrow \pi^+ X)}{d x_F d p_\perp^2}
+ E \frac{d^2 \sigma(pp \rightarrow \pi^- X)}{d x_F d p_\perp^2} \right]
\end{eqnarray}
for the invariant cross sections. The differential cross sections were 
parameterized by
\begin{eqnarray}
E \frac{d^2 \sigma(pp \rightarrow \pi^\pm X)}{d x_F d p_\perp^2}
= A_\pm \exp( B_\pm |x_F| + C_\pm x_F^2) \exp( D_\pm p_\perp + E_\pm p_\perp^2)
\end{eqnarray}
in \cite{Ammosov:1976zj}. Here $x_F = p_L/p_L^{\rm max}$ denotes Feynman-$x$ and 
$p_\perp$ the transverse momentum. The normalizations obey $A_+/A_- = 2.16 \pm 
0.24$ and the other parameters are  
\restylefloat{table}
\begin{table}[H]  
\small
\begin{center}
\renewcommand{\arraystretch}{1.5}
\begin{tabular}{|c|c|c|}
\hline
&  $+$ & $-$ \\
\hline
B  & $-5.21 \pm 1.03$ & $-9.51 \pm 0.21$ \\
C  & $-1.80 \pm 2.62$ & $+2.14 \pm 1.74$ \\
D  & $-1.80 \pm 0.31$ & $-1.22 \pm 0.38$ \\
E  & $-4.26 \pm 0.48$ & $-4.44 \pm 0.58$ \\
\hline
\end{tabular}
\renewcommand{\arraystretch}{1.5}
\end{center}
\end{table}

\begin{center}
\begin{figure}[H] 
\begin{center}
\epsfig{figure=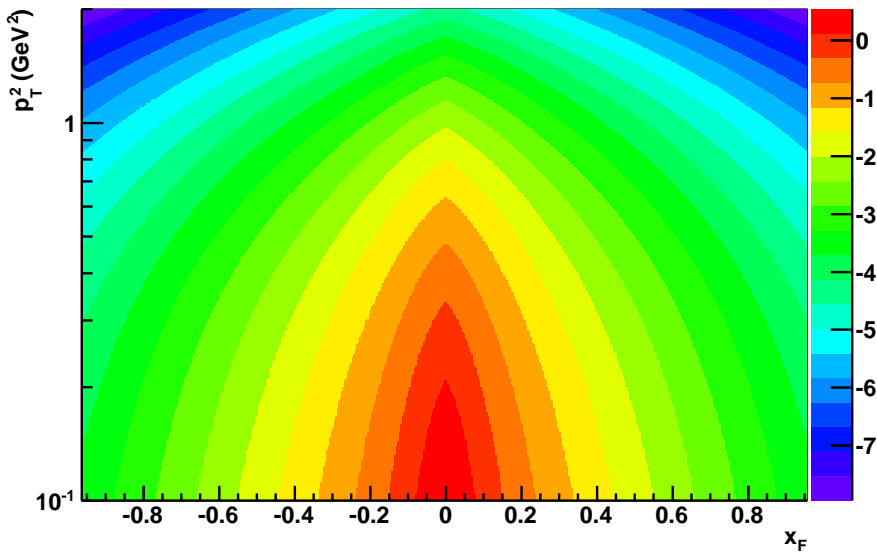,angle=0,width=0.7\linewidth} 
\end{center}
\caption[]{
\label{FIG:1}
\small
$p_\perp^2$ vs. $x_F$ distribution of the $\pi^0$ production spectrum in $p Fe$
collisions at $p = 68.6$~GeV. The color scale is in $\log_{10}$.}
\end{figure}
\end{center}

\begin{center}
\begin{figure}[H] 
\epsfig{figure=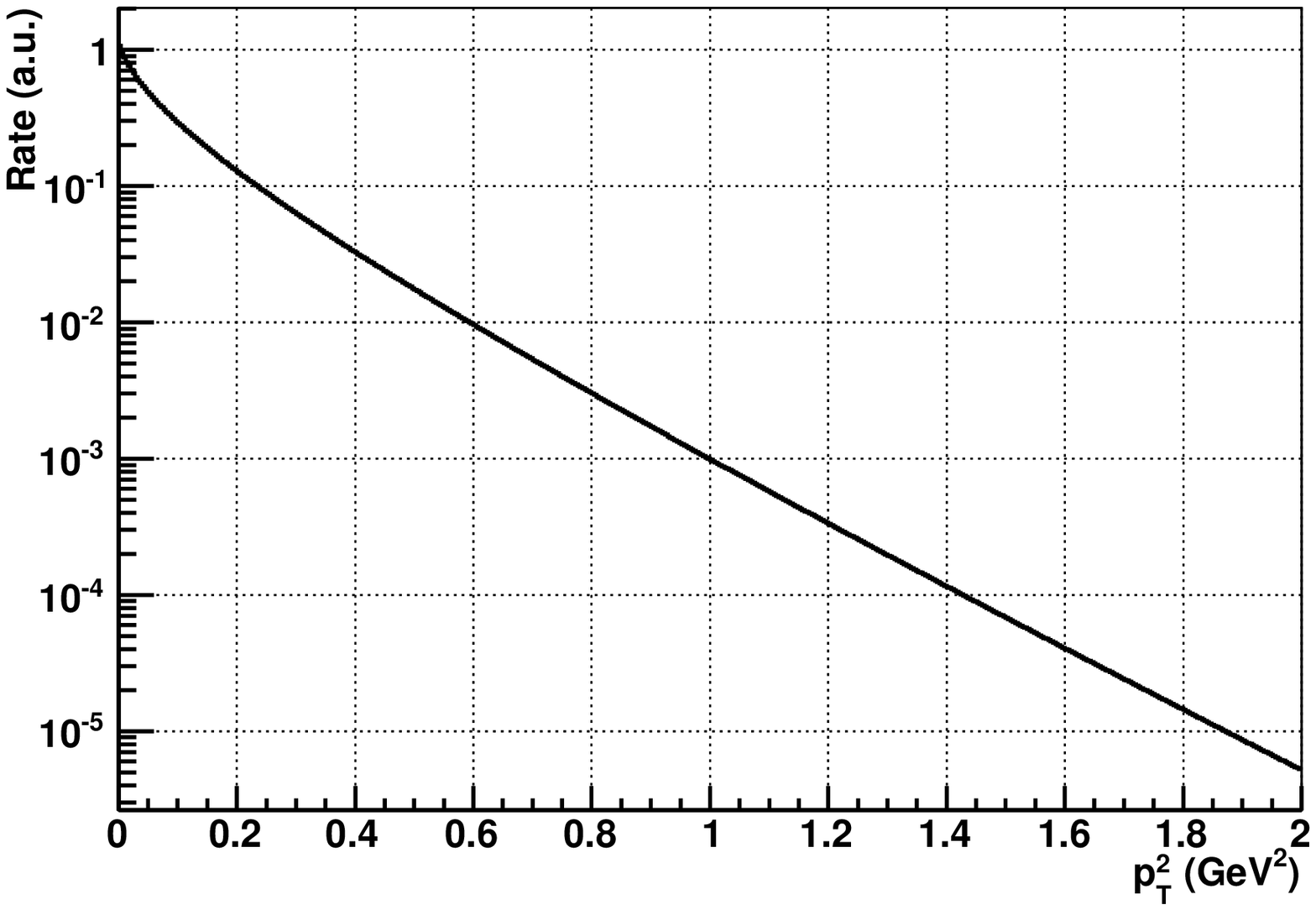,angle=0,width=0.47\linewidth} \hspace*{2mm}
\epsfig{figure=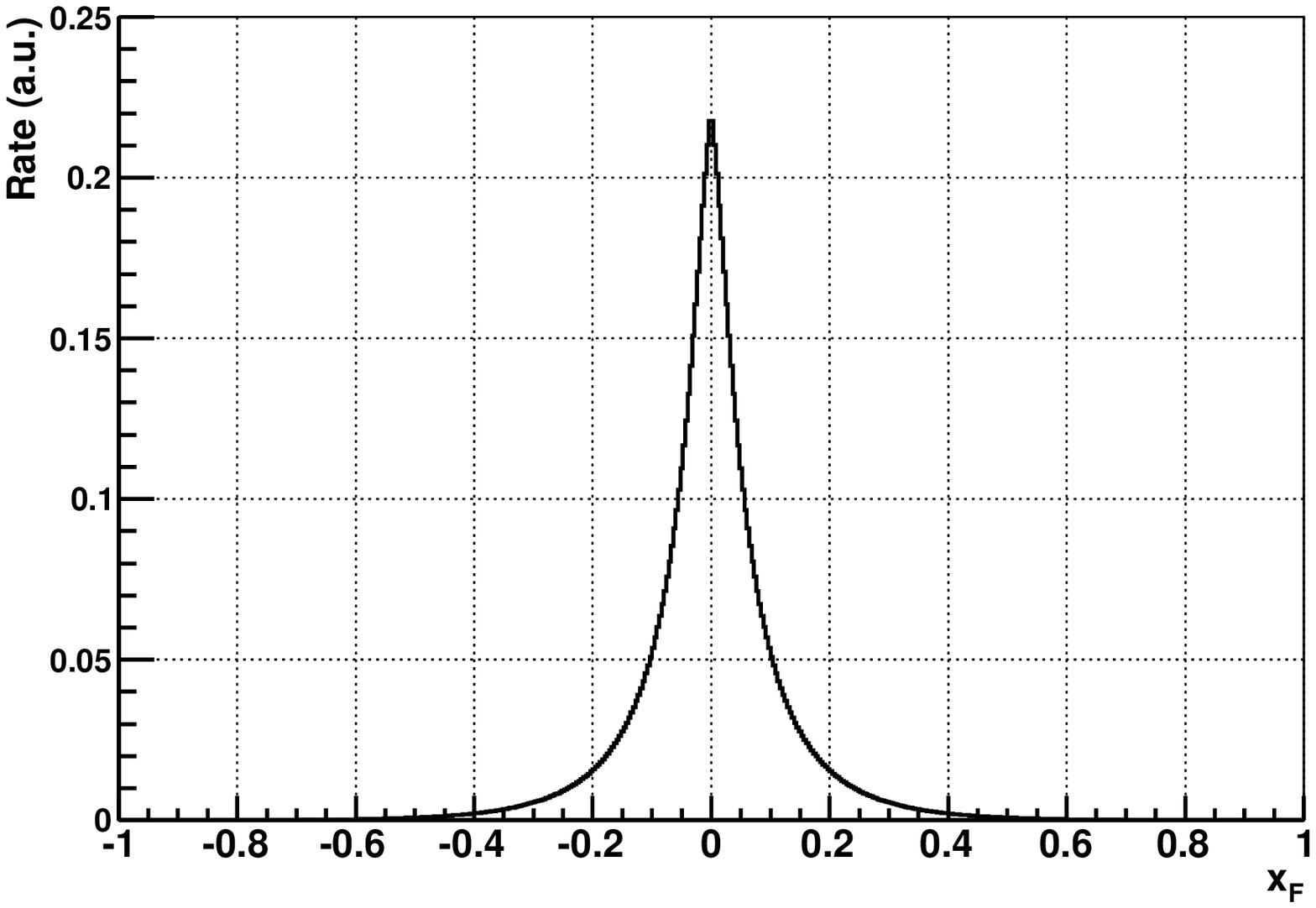,angle=0,width=0.47\linewidth} 
\caption[]{
\label{FIG:2}
\small
The $p_\perp^2$ and  $x_F$ distribution of the $\pi^0$ production spectrum in $p 
Fe$
collisions at $p = 68.6$~GeV.}
\end{figure}
\end{center}

\vspace*{-5mm}\noindent
Here we refer for the $p_\perp$ distribution to the values of $\langle p_\perp \rangle = 0.2$~GeV.
The resulting distributions in $p_\perp^2$ and $x_F$ are shown in
Figures~\ref{FIG:1},\ref{FIG:2}.

\vspace*{-5mm}\noindent
\restylefloat{figure}
The actual beam dump experiment used an iron target. The inclusive iron-proton 
cross section is related to the proton-proton cross section from (\ref{pp-tot}) by
\begin{equation}
\sigma(pFe \rightarrow \pi^0 X) = A^{\alpha(x_F)}\sigma(pp \rightarrow \pi^0 X)~,
\end{equation}  
with $A = 56$. The 
$A^{\alpha(x_F)}$ dependence of the $\pi^0$--production cross section on nuclei
as of iron can be parameterized by 

\begin{eqnarray}
\langle \alpha(x_F)\rangle \simeq 0.55~,
\end{eqnarray}
based on the compilation given in \cite{Barton:1982dg}. The inclusive cross section
at 69 GeV was measured with  \cite{Boratav:1976wx}
\begin{eqnarray}
\sigma(pp \rightarrow \pi^0 X) =  74 \pm 1.0 ~{\rm mb}~.
\end{eqnarray}
\restylefloat{figure}

\begin{center}
\begin{figure}[H] 
\begin{center}
\epsfig{figure=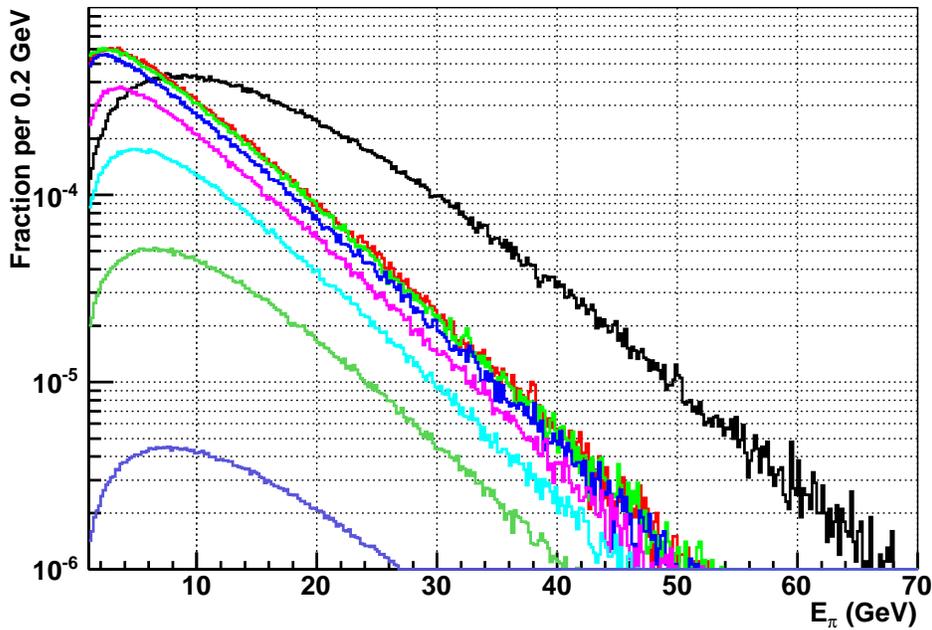,angle=0,width=0.8\linewidth} 
\end{center}
\caption[]{
\small
Fraction of produced $\gamma'$ particles which reach the detector
as function of their energy in the laboratory frame. Colors:
$\gamma'$ with masses between 0 and 120~MeV in steps of 20~MeV (from top to
bottom). The black line shows the corresponding $\pi^0$ for comparison.}
\label{FIG:3}
\end{figure}
\end{center}

\vspace*{-7mm}
We generate the $\gamma'$ particle production in the $pp$ center-of-momentum system 
and boost to the laboratory system then.  We investigate the mass range of $\gamma'$ 
particles above the 2-electron threshold of 1.022 MeV up to 
$m_{\pi^0}$. A cut $(p_\perp/p_L)_{\rm lab} < \tan(\Theta_{max})$ is applied 
to ensure that the produced particle reaches the fiducial volume
of the detector (see below). The effect of this cut is illustrated in 
Figure~\ref{FIG:3} which shows the fraction of particles which pass 
the fiducial volume cut with respect to all produced particles per energy bin. 
Results for $\pi^0$ are given as well as for
$\gamma'$ of various masses. The kinematic factor 
$(1-m_{\gamma'}^2/m_{\pi^0}^2)^3$ from (\ref{pp-tot}) is included for the 
$\gamma'$.
\restylefloat{figure}
The only relevant decay channel in the considered mass range is
$\gamma' \rightarrow e^+ e^-$.
The inverse live time of the $\gamma'$ particle is then given by 
\cite{Essig:2010gu} 
\begin{eqnarray}
\tau(\gamma')^{-1} = \frac{1}{3} \alpha_{QED} m_{\gamma'} \epsilon^2 \sqrt{1 - 
\frac{4 m_e^2}{m_{\gamma'}^2}}\left(1 + 
\frac{2 m_e^2}{m_{\gamma'}^2}\right)~.
\end{eqnarray}
Potential electron-pairs from $\gamma'$ decay manifest as electromagnetic showers
in the detector used.
The mass and coupling limits on the $\gamma'$ bosons given 
below are derived from the observed rate of these  showers over the 
background, cf. \cite{Blumlein:1990ay}.

\section{The Experimental Setup and Data Taking}

\vspace*{1mm}
\noindent
The beam dump experiment was carried out at the U70 accelerator at IHEP Serpukhov
during a three months exposure in 1989. Data have been taken
with the $\nu$-CAL I experiment, a neutrino detector. All technical 
details of this experiment have been described in \cite{Blumlein:1990ay}
and a detailed description of the detector was given in \cite{Barabash:2002zd}.
Here we only summarize the key numbers which are crucial for the present analysis.
 
The target part of the detector is used as a fiducial volume to detect the decays 
of the $\gamma'$. It has a modular structure and consists of 36 identical modules
along the beam direction. Each of the modules is composed of a 5~cm thick aluminum 
plate 
and a pair of drift chambers to allow for three dimensional tracking of charged 
particles.

For the beam dump experiment a fiducial volume of 30 modules with a total length of $l=23$~m 
is chosen, starting with the fourth module at a distance of 
$l_{\rm dump}=64$~m down-stream of the beam dump. The lateral extension of the 
fiducial volume
is $2.6 \times 2.6$~$\mbox{m}^2$. In the following we use conservatively a slightly smaller
fiducial volume, defined as a cone pointing to the beam dump with a ground circle of 
2.6~m in diameter at the end of the fiducial volume, i.e. at a distance of 87~m 
from the dump.
This leads to the following simple fiducial volume cut:
\begin{equation}
(p_\perp/p_L)_{lab} < 1.3/87 = 0.015~.
\end{equation}
For particles which traverse the fiducial volume, the decay probability 
$w_{\rm dec}$ for decays $\gamma' \rightarrow e^+ e^-$ is then given by
\begin{eqnarray}
w_{\rm dec} = 
\exp\left[- \frac{l_{\rm dump} }{c \tau(\gamma')} 
\frac{m_{\gamma'}}{p}\right] 
\left[1 - \exp\left(\frac{l }{c \tau(\gamma')}\frac{m_{\gamma'}}{p}
\right)\right]~,
\end{eqnarray}
with $c$ the velocity of light, $m_{\gamma'}$ and $p$ are
the mass and momentum of the $\gamma'$.

During the three months exposure time in 1989 $1.71\times 10^{18}$ protons on target had been 
accumulated~\cite{Blumlein:1990ay}. The signature of event candidates from 
$\gamma'\rightarrow e^+e^-$ is a single 
electromagnetic shower in beam direction. This signature is identical to the one from 
the axion or light Higgs particle decay search which was performed 
in~\cite{Blumlein:1990ay}. 
For energy deposits above 3~GeV in the detector the reconstruction 
code  can distinguish electromagnetic from hadronic showers very well. Therefore 
the
following final cuts for the selection of isolated electromagnetic showers had been 
chosen~\cite{Blumlein:1990ay}~:
\begin{itemize}
\item A minimal electromagnetic shower energy of $E_{\rm elm} > 3$~GeV; 
\item A maximal hadronic shower energy of $E_{\rm had} < 1.5$~GeV;
\item A maximal angle with respect to the beam direction of $\Theta_{\rm elm} < 
0.05$~rad.
\end{itemize}
From the total data sample of 3880 reconstructed events, 5 pass these cuts. Background 
estimates from the simulation of $\nu_\mu$ and $\nu_e$ interactions in the detector 
account for 3.5~events. The Poisson probability to observe 5 or less events for an 
expectation of 3.5~events is 86\%. Data are therefore compatible with the simulated 
background from conventional neutrino interactions.
\section{\boldmath Search for decays $\gamma'\rightarrow e^+e^-$}

\begin{center}
\begin{figure}[H] 
\begin{center}
\epsfig{figure=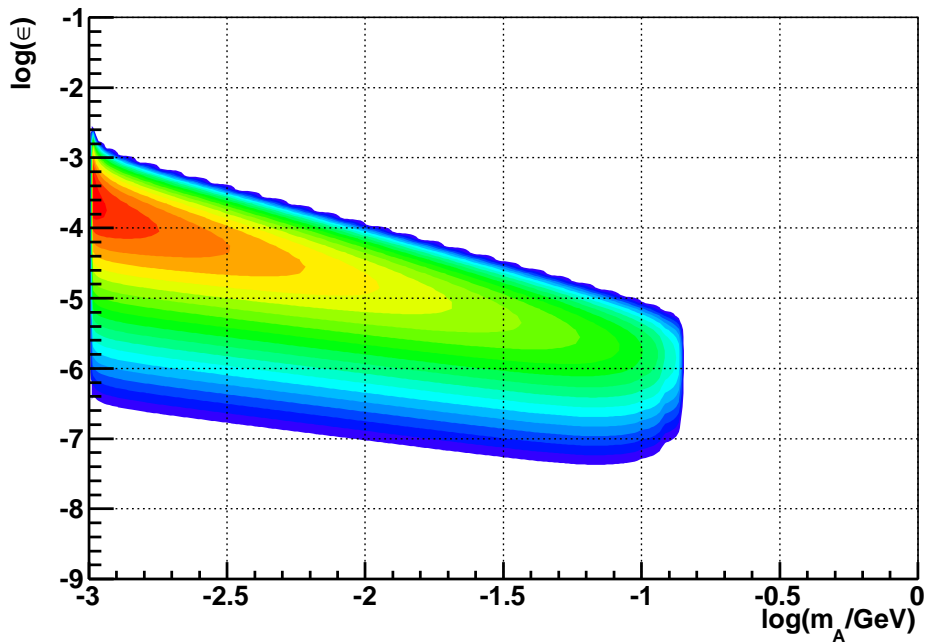,angle=0,width=0.8\linewidth} \hspace*{2mm}
\end{center}
\caption[]{
\small
Detected $\gamma'$ events. Color scale in $\log_{10}$ from $10^8$ events (red) to one event
(dark blue).}
\label{FIG:4}
\end{figure}
\end{center}

\vspace*{1mm}
\noindent
Signals from $\gamma'\rightarrow e^+e^-$ decays  pass the cuts  mentioned above  
with an energy independent efficiency of $\varepsilon=70\%$ if the true energy 
of the decaying particle 
is above 3~GeV. The total number of expected signal events can therefore be calculated as
\begin{equation}
N_{\rm sig} = N_{\rm tot}\times
\frac{\sigma(pFe \rightarrow \gamma' X)_{\rm forward}}{\sigma(pp \rightarrow X)}
\times w_{\rm dec}\times\varepsilon~,
\end{equation}
with $N_{\rm tot}$ the total number of protons on target during the exposure time.
The index `forward' indicates the application of the fiducial volume cut.
The dependence of $N_{\rm sig}$ on $m_{\gamma'}$ and $\epsilon$ is shown in 
Figure~\ref{FIG:4}.
\restylefloat{figure}
For $m_{\gamma'}\approx 2$~MeV and $\epsilon\approx 2\cdot 10^{-4}$ the decay probability
reaches its maximum of 11\% and more than $10^8$ signal events would be expected in the
detector. For larger $\epsilon$ the decay length decreases exponentially and at some point
all $\gamma'$ particles decay before reaching the detector. For smaller values of 
$\epsilon$ the decay length
increases. At $\epsilon< 10^{-6}$ most of the particles pass the detector
without decaying. For even smaller values of $\epsilon$ the event rate of 
both production and decay probability decrease proportional to $\epsilon^2$. 
The dependence of the iso-event number lines on $m_{\gamma'}$ is governed by the 
boost factor $p/m_{\gamma'}$. The sensitivity is
kinematics limited to values $m_{\gamma'}<m_{\pi^0}$.

\begin{center}
\begin{figure}[H] 
\begin{center}
\epsfig{figure=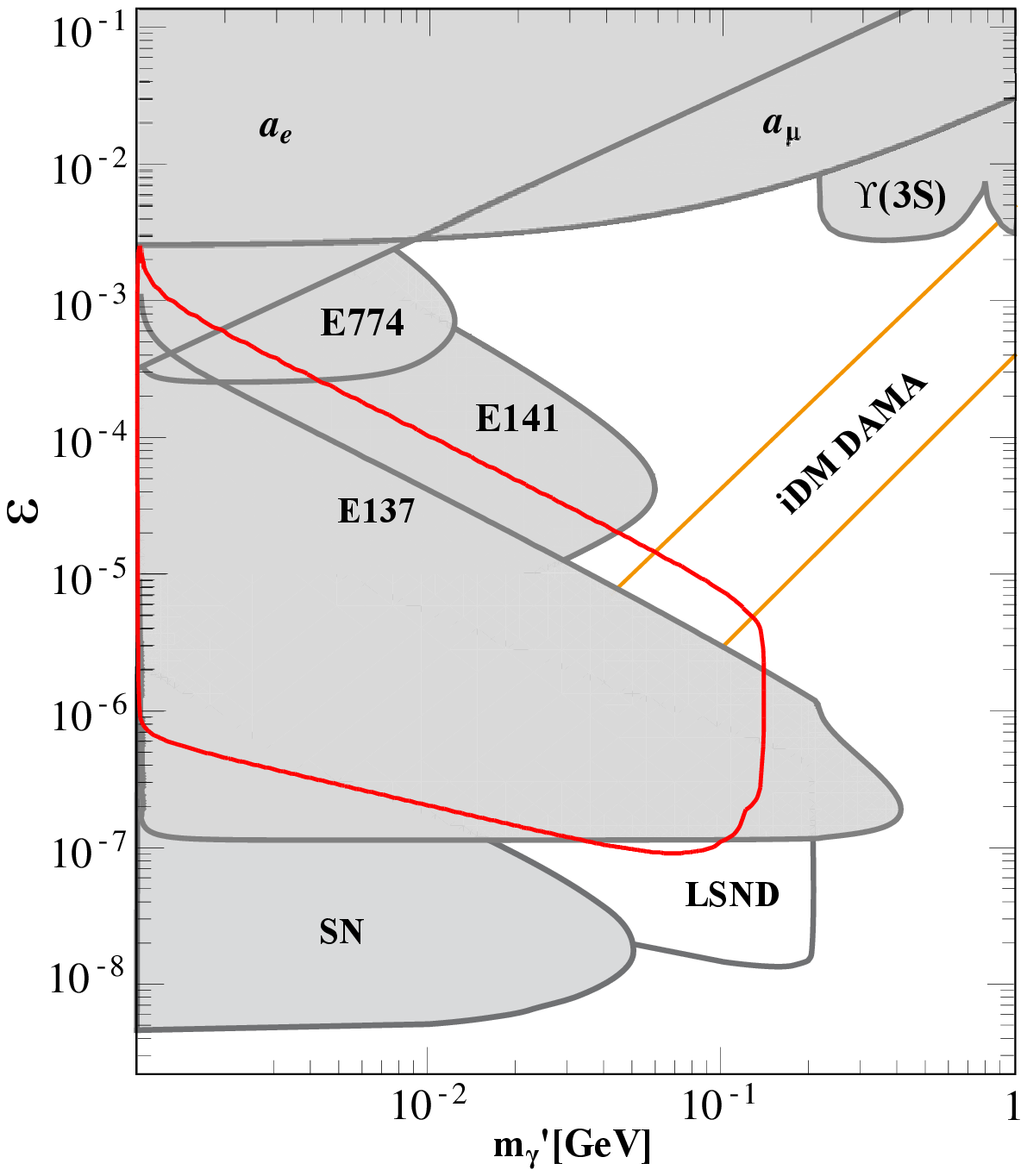,angle=0,width=0.8\linewidth} 
\end{center}
\caption[]{
\small
Comparison of the present exclusion bounds (red line) with other limits from the 
measurement of the anomalous magnetic moments $a_e$ and 
$a_\mu$~\cite{Pospelov:2008xx}, $\Upsilon(3S)$ decay \cite{UPS}, the beam dump 
experiments E137, E141, E774 \cite{E137,E141,E774}, and supernovae 
cooling~\cite{Turner:1987by,Bjorken:2009mm}. We indicate the prospects for 
LSND 
\cite{Essig:2010gu,LSND} (open grey-bounded area), and  the DAMA/LIBRA region (open orange 
bounded area) \cite{DAMA}. The limits for $\epsilon > 10^{-7}$ have been taken 
from Ref.~\cite{Andreas:2010tp}.} 
\label{FIG:5} \end{figure}
\end{center}

For 10.6 expected events the Poisson probability to observe 
5 or less events is less than 5\%.
With a background expectation of 3.5~events we can therefore exclude 
a signal contribution of 7.1~events
at 95\% C.L. The corresponding exclusion region is shown as the red line 
in comparison with limits from other experiments in Figure~\ref{FIG:5}, 
see also \cite{Bjorken:2009mm,Essig:2010gu,Andreas:2010tp}.
\restylefloat{figure}
At large values of $\epsilon$ studies of the anomalous magnetic moments of the muon and 
electron~\cite{Pospelov:2008xx} and 
of rare decays of heavy mesons~\cite{UPS} put stringent limits. For 
$10^{-3} < \epsilon < 10^{-7}$ beam dump experiments~\cite{E137,E141,E774} give the
best sensitivity. For even smaller values of $\epsilon$ limits can be derived
by studying the dynamics of supernovae cooling~\cite{Turner:1987by}. 
For completeness, the prospects for the sensitivity of a reanalysis of LSND 
data~\cite{LSND} is also shown, as proposed in~\cite{Essig:2010gu}.

The present analysis
is sensitive to the region $10^{-3} < \epsilon < 10^{-7}$ and the exclusion region 
largely
overlaps with the one from E137. However, we are sensitive to larger values of $\epsilon$
for the same values of $m_{\gamma'}$. In particular, a new region around 
$\epsilon\approx 10^{-5}$
and $m_{\gamma'}\approx 50$~MeV is explored here. This region is part of a band (
shown in
orange) which correlates $\epsilon$ and $m_{\gamma'}$ within certain 
supersymmetric 
theories~\cite{Dienes:1996zr}. Within these models heavy dark matter candidate particles 
pick up an $U(1)'$ charge and can scatter elastically through $\gamma'$ exchange. 
This would allow to explain the annual modulation of the DAMA/LIBRA 
experiment~\cite{DAMA}.
For a more extensive discussion, see~\cite{Bjorken:2009mm}. 
\section{Conclusions}

\vspace*{1mm}
\noindent
We have re-analyzed proton beam dump data taken at the U70 accelerator at IHEP
Serpukhov with the $\nu$-calorimeter I experiment in 1989 to set mass and coupling limits for
dark gauge forces, searching for electromagnetic signatures according to the decay 
$\gamma' \rightarrow e^+ e^-$. The analysis extends the region excluded by former experiments
in the mass region $m_{\gamma'} \in [0.03~{\rm GeV}, m_{\pi_0} ]$ towards larger 
values in the mixing 
parameter $\epsilon \in [2 \times 10^{-6}, 2 \times 10^{-5}]$. A lower part of the 
anticipated DAMA/LIBRA region is 
excluded. At lower values of
$\epsilon \approx 10^{-7}$ a smaller region of masses in the range $m_{\gamma'} \in [0.03, 0.1 ]$~GeV 
is excluded beyond the bounds given by E137 \cite{E137}. In future experiments signals from dark gauge forces 
will be searched for in the yet unexplored regions shown in Figure~\ref{FIG:5}, see e.g. Ref.~\cite{Andreas:2010tp} 
for proposals.

\vspace*{4mm}
\noindent
{\bf Acknowledgment.}~
{We would like to thank our former colleagues of the $\nu$-Cal I experiment for
collaboration during the time 1983--1990. For discussions we thank M. Walter 
and A. Ringwald. This paper has been supported in part by 
DFG Sonderforschungsbereich Transregio 9, Computergest\"utzte Theoretische 
Teilchenphysik and EU Network {\sf LHCPHENOnet}
PITN-GA-2010-264564.}


\end{document}